\documentclass{article}
\usepackage{ijcai19}

\usepackage{times}
\usepackage{soul}
\usepackage{url}
\usepackage[hidelinks]{hyperref}
\usepackage[utf8]{inputenc}
\usepackage[small]{caption}
\usepackage{graphicx}
\usepackage{amsmath}
\usepackage{booktabs}
\usepackage{algorithm}
\usepackage{algorithmic}
\usepackage{comment}
\usepackage{mathtools}
\usepackage{amsmath, amsthm, amssymb, amsfonts}
\usepackage{comment} 
\usepackage{subfig}
\usepackage{mathrsfs}
\usepackage{textcomp}
\usepackage{xcolor}
\usepackage{booktabs}
\usepackage{multirow}

\title{Multi-agent Attentional Activity Recognition}

\author{
Kaixuan Chen$^1$
\and
Lina Yao$^1$\and
Dalin Zhang$^{1}$\and
Bin Guo$^{2}$\And
Zhiwen Yu$^2$
\affiliations
$^1$University of New South Wales\\
$^2$Northwestern Polytechnical University
\emails
\{kaixuan.chen@student., lina.yao@\}unsw.edu.au,
}
\begin{document}

\maketitle

\begin{abstract}

Multi-modality is an important feature of sensor based activity recognition. In this work, we consider two inherent characteristics of human activities, the spatially-temporally varying salience of features and the relations between activities and corresponding body part motions. Based on these, we propose a multi-agent spatial-temporal attention model. The spatial-temporal attention mechanism helps intelligently select informative modalities and their active periods. And the multiple agents in the proposed model represent activities with collective motions across body parts by independently selecting modalities associated with single motions. With a joint recognition goal, the agents share gained information and coordinate their selection policies to learn the optimal recognition model.
The experimental results on four real-world datasets demonstrate that the proposed model outperforms the state-of-the-art methods.

% and propose a globally attentive selection strategy to select salient features spatially and temporally. Moreover, considering that activities comprises motions from different body parts, the proposed model employs multi-agent to represent the combination of the motions. We evaluate our model in four real-world datasets. The experiment results demonstrate that the proposed model outperforms the state-of-the-art methods.
\end{abstract}

\section{Introduction}

The ability to identify human activities via on-body sensors has been of interest to the healthcare community \cite{anguita2013public}, the entertainment \cite{freedman2018roles} and fitness \cite{guo2017fitcoach} community. 
Some works of Human Activity Recognition (HAR) are based on hand-crafted features for statistical machine learning models \cite{lara2013survey}. Until recently, deep learning has experienced massive success in modeling high-level abstractions from complex data \cite{pouyanfar2018survey}, and there is a growing interest in developing deep learning for HAR \cite{hammerla16deep}. 
Despite this, these methods still lack sufficient justification when being applied to HAR.
In this work, we consider two inherent characteristics of human activities and exploit them to improve the recognition performance.

The first characteristic of human activities is the spatially-temporally varying salience of features.
% the heterogeneous sensory data and 
Human activities can be represented as a sequence of multi-modal sensory data. The modalities include acceleration, angular velocity and magnetism from different positions of testers' bodies, such as chests, arms and ankles. However, only a part of modalities from specific positions are informative for recognizing certain activities \cite{wang2017modeling}. Irrelevant modalities often influence the recognition and undermine the performance. For instance, identifying lying mainly relies on people's orientations (magnetism), and going upstairs can be easily distinguished by upward acceleration from arms and ankles. In addition, the significance of modalities changes over time. Intuitively, the modalities are only important when the body parts are actively participating in the activities.
Therefore, we propose a spatial-temporal attention method to select salient modalities and their active periods that are indicative of the true activity.
Attention has been proposed as a sequential decision task in earlier works \cite{denil2012learning,mnih2014recurrent}.
This mechanism has been applied to sensor based HAR in recent years.
\cite{chen2018interpretable} and \cite{zhang2018multi} transform the sensory sequences into 3-D activity data by replicating and permuting the input data, and they propose to attentionally keep a focal zone for classification. However, these methods heavily rely on data pre-processing, and the replication increases the computation complexity unnecessarily.
Also, these methods do not take the temporally-varying salience of modalities into account. In contrast, the proposed spatial-temporal attention approach directly selects informative modalities and their active time that are relevant to classification from raw data. The experiment results shows that our model makes HAR more explainable.

\begin{figure*}[htbp]
\centering
\includegraphics[width=7.0in]{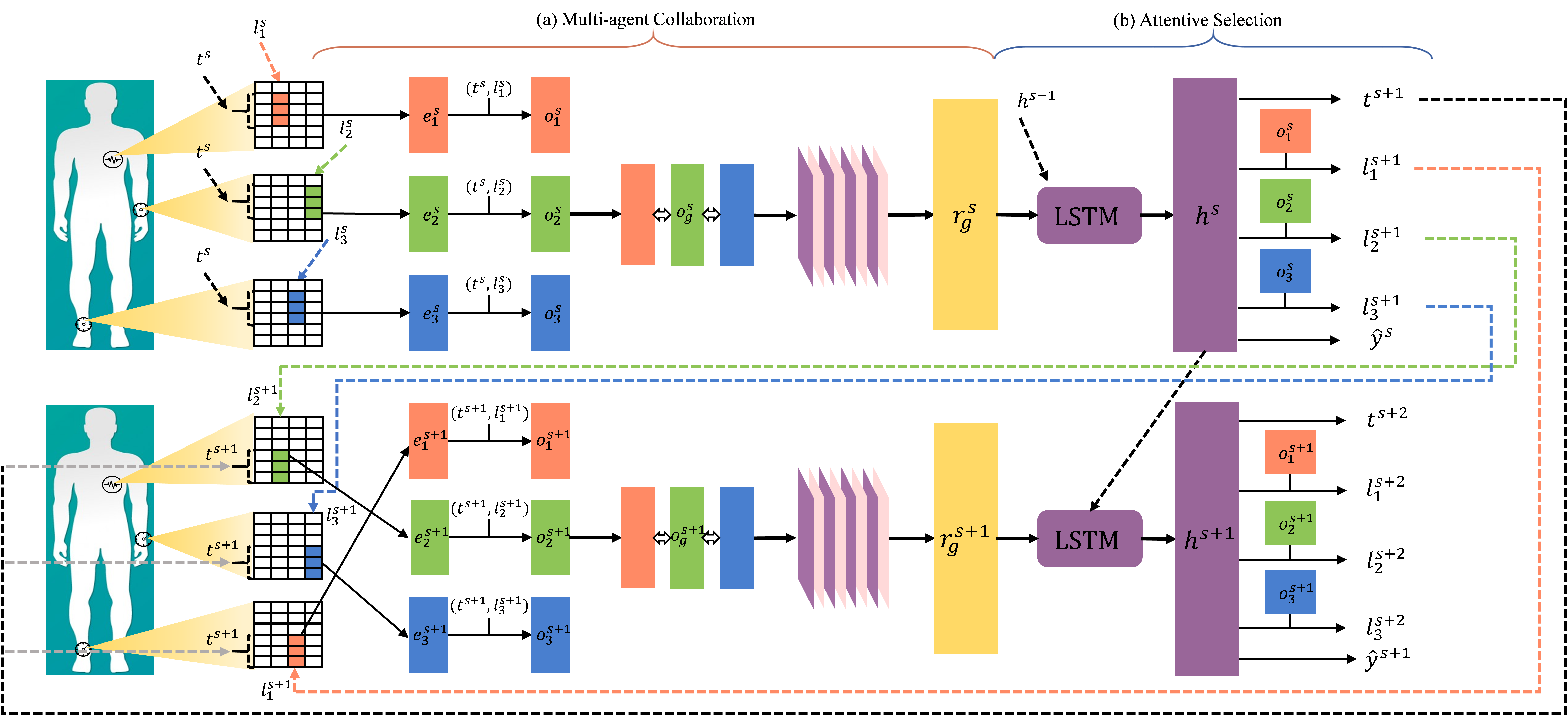} 
\caption{The overview of the proposed model. At each step $s$, three agents $a_1, a_2, a_3$ individually select modalities and obtain observations $o_1^s, o_2^s, o_3^s$ from the input $\textbf{x}$ at $(t^s, l_1^s)$, $(t^s, l_2^s)$ and $(t^s, l_3^s)$. 
The agents then exchange and process the gained information to get the representation $r_g^s$ of the shared observation. And they decide the next locations again. Based on a sequence of observations after an episode, the agents jointly make the classification. Red, green and blue denote the workflows that are associated with $a_1, a_2, a_3$, respectively. Other colors denote the shared information and its representations.}
\label{fig:overview}
\end{figure*}

The second characteristic of human activities considered in this paper is activities are portrayed by motions on several body parts collectively. 
For instance, running can be seen as a combination of arm and ankle motions. Some works like \cite{radu2018multimodal,yang2015deep} are committed to fusing multi-modal sensory data for time-series HAR, but they only fuse the information of local modalities from the same positions. These methods, as well as the existing attention based methods \cite{chen2018interpretable,zhang2018multi}, are limited in capturing the global interconnections across different body parts. 
To fill this gap, we propose a multi-agent reinforcement learning approach.  
We simplify activity recognition by dividing the activities into sub-motions with which an independent intelligent agent is associated and by coordinating the agents' actions. 
These agents select informative modalities independently based on both their local observations and the information shared by each other. 
Each agent can individually learn an efficient selection policy by trial-and-error. 
After a sequence of selections and information exchanges, a joint decision on recognition is made.
The selection policies are incrementally coordinated during training since the agents share a common goal which is to minimize the loss caused by false recognition. 

The key contributions of this research are as follows:
\begin{itemize}
\item We propose a spatial-temporal attention method for temporal sensory data, which considers the spatially-temporally varying salience of features, and allows the model to focus on the informative modalities that are only collected in their active periods.
\item We propose a multi-agent collaboration method. The agents represent activities with collective motions by independently selecting modalities associated with single motions and sharing observations.
The whole model can be optimized by coordinating the agents' selection policies with the joint recognition goal.
\item We evaluate the proposed model on four datasets. The comprehensive experiment results demonstrate the superiority of our model to the state-of-the-art approaches.
\end{itemize}

\section{The Proposed Method}

\subsection{Problem Statement}
We now detail the human activity recognition problem on multi-modal sensory data.
Each input sample $(\textbf{x}, y)$ consists of a 2-d vector $\textbf{x}$ and an activity label $y$. Let $\textbf{x}$ = $[\textbf{x}^0, \textbf{x}^1, ...\textbf{x}^K]$ where $K$ denotes the time window length and $\textbf{x}^i$ denotes the sensory vector collected at the point $i$ in time.
$\textbf{x}^i$ is the combination of multi-modal sensory data collected from testers' different body positions such as chests, arms and ankles. 
Suppose that
$\textbf{x}^i$ = $(\textbf{m}^i_1, \textbf{m}^i_2, ... \textbf{m}^i_N)$ = $(x^i_0, x^i_1, ...x^i_P)$, where $\textbf{m}$ denotes data collected from each position, $N$ denotes the number of positions (in our datasets, $N = 3$), and $P$ denotes the number of values per vector.
Therefore, $\textbf{x}\in R^{K\times P}$ and $y\in [1, ..., C]$. $C$ represents the number of activity classes. The goal of the proposed model is to predict the activity $y$.

\subsection{Model Structure}
% In this work we propose a multi-agent spatial-temporal attention model for HAR.
The overview of the model structure is shown in Figure~\ref{fig:overview}. At each step $s$, the agents select an active period together and individually select informative modalities from the input $\textbf{x}$. 
These agents share their information and independently decide where to ``look at" at the next step. The locations are determined spatially and temporally in terms of modalities and time.  After several steps, the final classification is jointly conducted by the agents based on a sequence of the observations.
Each agent can incrementally learn an efficient decision policy over episodes. But by having the same goal, which is to jointly minimize the recognition loss, they collaborate with each other and learn to align their behaviors such that it achieves their common goal.
\subsubsection{Multi-agent Collaboration.}
In this work we simplify activity recognition by dividing the activities into sub-motions and require each agent select informative modalities that are associated with one motion.
Suppose that we employ $H$ agents $a_1, a_2,... a_H$ (we assume $H=3$ in this paper for simplicity). The workflows of $a_1, a_2, a_3$ are shown in red, green and blue in Figure~\ref{fig:overview}. 
% At each step $s$, the agents make individual observations. They also decide where to make the next observations separately.
% 

At each step $s$, each agent locally observes a small patch of $\textbf{x}$, which includes information of a specific modality from a motion in its active period. 
Let the observations be $e_1^s, e_2^s, e_3^s$ as Figure~\ref{fig:overview} shows. They are extracted from $\textbf{x}$ at the locations $(t^s, l_1^s)$, $(t^s, l_2^s)$ and $(t^s, l_3^s)$, respectively, where $t$ denotes the selected active period and $l$ denote the location of a modality in the input $\textbf{x}$. The model encodes the region around $(t^s, l_i^s)$ ($i \in \{1,2,3\}$) with high resolution but uses a progressively lower resolution for points further from $(t^s, l_i^s)$ in order to remove noises and avoid information loss in \cite{zontak2013separating}. 
We then further encode the observations into higher level representations.
With regard to each agent $a_i$ ($i \in \{1,2,3\}$), the observation
$e_i^s$ and the location $(t^s, l_i^s)$ are linear transformed independently, parameterized by $\theta_e$ and $\theta_{tl}$, respectively. Next, the summation of these two parts is further transformed with another linear layer parameterized by $\theta_o$ The whole process can be summarized as the following equation:
\begin{eqnarray}
\label{eqn:partial}
o_i^s &=& f_o(e_i^s, t^s, l_i^s;\theta_e, \theta_{tl}, \theta_o) \nonumber \\
&=& L(L(e_i^s) + L(concat(t^s, l_i^s))) \;\;i \in \{1,2,3\},
\end{eqnarray}
where $L(\bullet)$ denotes a linear transformation and $concat(t^s, l_i^s)$  represents the concatenation of $t^s$ and $l_i^s$. Each linear layer is followed by a rectified linear unit (ReLU) activation. Therefore, $o_i^s$ contains information from "what" ($\rho (C^f, l^f_t)$), "where" ($l^f_t$) and "when".

Making multiple observations not only avoids the system processing the whole data at a time but also maximally prevents the information loss from only selecting one region of data. Furthermore, multiple agents make observations individually so that they can represent activities with the collective modalities from different motions.
The model can explore various combinations of modalities to recognize activities during learning. 

Then we are interested in the collaborative setting where the agents communicate with each other and share the observations they make. So we get the shared observation $o_g^s$ by concatenate $o_1^s, o_2^s, o_3^s$ together.
\begin{equation}
\label{eqn:global}
o_g^s = concat(o_1^s, o_2^s, o_3^s),
\end{equation}
so that $o_g^s$ contains all the information observed by three agents. A convolutional network is further applied to process $o_g^s$ and extract the informative spatial relations. The output is then reshaped to be the representation $r_g^s$.
\begin{equation}
\label{eqn:conv}
r_g^s = f_c(o_g^s;\theta_c) = reshape(Conv(o_g^s))
\end{equation}
And $r_g^s$ represents the activity to be identified with multiple modalities selected from motions on different body positions.

\subsubsection{Attentive Selection.}
In this section, the details about how to select modalities and active period attentively are introduced. 
We first introduce the episodes in this work.
The agents incrementally learn the attentive selection policies over episodes.
% In the reinforcement learning setting, an episode is one complete play of the agent interacting with the environment \cite{deloach2006agent}. 
In each episode, following the bottom-up processes, the model attentively selects data regions and integrates the observations over time to generate dynamic representations, in order to determine effective selections and maximize the rewards, i.e., minimize the loss. Based on this, LSTM is appropriate to build an episode as it incrementally combines information from time steps to obtain final results. As can be seen in Figure~\ref{fig:overview}, at each step $s$, the LSTM module receives the representation $r_g^s$ and the previous hidden state $h^{s-1}$ as the inputs. Parameterized by $\theta_h$, it outputs the current hidden state $h^s$:
\begin{equation}
\label{eqn:hidden}
h^s = f_h(r_g^s, h^{s-1};\theta _h)
\end{equation}

Now we introduce the selection module.
% The selection module is a principal component of the proposed model. 
The agents are supposed to select salient modalities and an active period at each step. To be specific, they need to select the locations where they make next observations.
Three agents control $l^{s+1}_1, l^{s+1}_1, l^{s+1}_3$ independently based on both the hidden state $h^s$ and their individual observations $o_1^s, o_2^s, o_3^s$ so that the individual decisions are made from the overall observation as well. $t^{s+1}$ is jointly decided based on $h^s$ only since it is a common selection.
The decisions are made by the agents' selection policies which are defined by Gaussian distribution stochastic process:
\begin{equation}
\label{eqn:l}
l^{s+1}_i \sim P(\cdot \mid f_l(h^s, o_i^s; \theta_{l_i})) \;\; i \in \{1,2,3\},
\end{equation}
and
\begin{equation}
\label{eqn:t}
t^{s+1} \sim P(\cdot \mid f_t(h^s; \theta_{t}))
\end{equation}
The purpose of stochastic selections is to explore more kinds of selection combinations such that the model can learn the best selections during training. 
% Note that during test, the actions are only decided by linear layers $f_l(h^s, o_i^s; \theta_{l_i})$ and $f_t(h^s; \theta_{t})$.
% \subsubsection{Reward}

To align the agents' selection policies, we assign the agents a common goal that correctly recognizing activities after a sequence of observations and selections. They together receive a positive reward if the recognition is correct. Therefore, at each step $s$, a prediction $\hat{y}^s$ is made by:
\begin{equation}
\label{eqn:y}
\hat{y}^s = f_y(h^s; \theta_y) = softmax(L(h^s))
\end{equation}
Usually, agents receive a reward $r$ after each step. But in our case, since only the classification in the last step $S$ is representative, the agents receive a delayed reward $R$ after each episode.
\begin{equation}
\label{eqn:R}
R=\left\{
\begin{aligned}
1 \;\; if \;\; \hat{y}^S = y \\
0 \;\; if\;\;  \hat{y}^S \neq y
\end{aligned}
\right.
\end{equation}
The target of optimization is to coordinate all the selection policies by maximizing the expected value of the reward $\Bar{R}$ after several episodes.

\subsection{Training and Optimization}

This model involves parameters that define the multi-agent collaboration and the attentive selection. The parameters $\Theta = \{\theta_e, \theta_{tl}, \theta_o, \theta_c, \theta_h, \theta_{l_i}, \theta_t, \theta_y\}$ ($i \in \{1,2,3\}$). The parameters for classification can be optimized by minimizing the cross-entropy loss:
\begin{equation}
\label{eqn:L_c}
    L_c = -\frac{1}{N}\sum_{n=1}^{N}\sum_{c=1}^{C}y_{n}(c)\log F_y(\textbf{x}),
\end{equation}
where $F_y$ is the overall function that outputs $\hat{y}$ given $\textbf{x}$. $C$ is the number of activity classes, and $y_{n}(c) = 1$ if the $n$-th sample belongs to the $c$-th class and $0$ otherwise.

However, selection policies that are mainly defined by $\theta_{l_i}$ ($i \in \{1,2,3\}$) and $\theta_t$ are expected to select a sequence of locations. The parameters are thus non-differentiable. 
In this view, we deploy a Partially Observable Markov Decision Process (POMDP) \cite{cai2009learning} to solve the optimization problem. Suppose $e^s = (e^s_1, e^s_2, e^s_3)$, $lt^s = (l^s_1, l^s_2, l^s_3, t^s)$,  We consider each episode as a trajectory $\tau = \{e^1, lt^1, y^1; e^2, lt^2, y^2; ..., e^S, lt^S, y^S\}$. Each trajectory represents one order of the observations, the locations and the predictions our agents make. After agents repeat $N$ episodes, we can obtain $\{\tau^1, \tau^2,...,\tau^N\}$, and each $\tau$ has a probability $p(\tau;\Theta)$ to be obtained. The probability depends on the selection policy $\Pi = (\pi_1, \pi_2, \pi_3)$ of the agents.

Our goal is to learn the best selection policy $\Pi$ that maximizes $\Bar{R}$. Specifically, $\Pi$ is decided by $\Theta$. Thus we need to find out the optimized $\Theta^\ast = \underset{\Theta}{\arg\max} [\Bar{R}]$. One common way is gradient ascent.

Generally, given a sample $x$ with reward $f(x)$ and probability $p(x)$, the gradient can be calculated as follows:
\begin{eqnarray}
\label{eqn:R rule}
\nabla_\theta E_x[f(x)] &=& \nabla_\theta\sum_{x}p(x)f(x) \nonumber \\
%&=& \sum_{x}\nabla_\theta p(x)f(x) \nonumber\\
&=& \sum_{x}p(x)\frac{\nabla_\theta p(x)}{p(x)}f(x) \nonumber\\
&=& \sum_{x}p(x)\nabla_\theta log p(x)f(x)      \nonumber \\
&=& E_x[f(x)\nabla_\theta log p(x)]
\end{eqnarray}
In our case, a trajectory $\tau$ can be seen as a sample, the probability of each sample is $p(\tau;\Theta)$, and the reward function $\bar{R} = E_{p(\tau;\Theta)}[R]$. We have the gradient:
\begin{equation}
\nabla_\Theta \bar{R} = E_{p(\tau;\Theta)}[R\nabla_\Theta log p(\tau;\Theta)]
\end{equation}
By considering the training problem as a POMDP and following the REINFORCE rule \cite{williams1992simple}:
\begin{equation}
\nabla_\Theta \bar{R} = E_{p(\tau;\Theta)}[R\sum_{s=1}^{S}\nabla_\Theta log \Pi(y|\tau_{1:s};\Theta)]
\end{equation}

Since we need several samples $\tau$ for one input $\textbf{x}$ to learn the best policy combination, we adopt Monte Carlo sampling which utilizes randomness to yield results that might be theoretically deterministic. Supposing $M$ is the number of Monte Carlo sampling copies, we duplicate the same input for $M$ times and average the prediction results. The $M$ copies generate $M$ subtly different results owing to the stochasticity, so we have:
\begin{equation}
\label{eqn:delta L_R}
\nabla_\Theta \bar{R} \thickapprox \frac{1}{M} \sum_{i=1}^{M}R^{(i)} \sum_{s=1}^{S} \nabla_\Theta log \Pi(y^{(i)}|\tau^{(i)}_{1:s};\Theta),
\end{equation}
where $M$ denotes the number of Monte Carlo samples, $i$ denotes the $i^{th}$ duplicated sample, and $y_i$ is the correct label for the $i^{th}$ sample. Therefore, the overall optimization can be summarized as maximizing $\bar{R}$ and minimizing Eq.~\ref{eqn:L_c}. The detailed procedure is shown in Algorithm~\ref{alg: 1}.

\begin{algorithm}[!t]
\caption{Training and Optimization}
\label{alg: 1}
\begin{algorithmic}[1]

\REQUIRE sensory matrix $\textbf{x}$, label $y$, \\the length of episodes $S$, \\the number of Monte Carlo samples $M$.
 
\ENSURE  parameters $\Theta$.

\STATE $\Theta = RandomInitialize()$

\WHILE{training}

\STATE duplicate $\textbf{x}$ for $M$ times 
\FOR{$i$ from $1$ to $M$}
\STATE $l^{1(i)}_1, l^{1(i)}_2, l^{1(i)}_3, t^{1(i)} = RandomInitialize()$
\FOR{$s$ from $1$ to $S$}
\STATE extract $e^{s(i)}_1, e^{s(i)}_2, e^{s(i)}_2$
\STATE $o^{s(i)}_1, o^{s(i)}_2, o^{s(i)}_3\leftarrow Eq.~\ref{eqn:partial}$
\STATE $o_g^{s(i)},r_g^{s(i)},h^{s(i)}\leftarrow Eq.~\ref{eqn:global},Eq.~\ref{eqn:conv},Eq.~\ref{eqn:hidden}$

\STATE $l^{s(i)}_1, l^{s(i)}_2, l^{s(i)}_3, t^{s(i)} \leftarrow Eq.~\ref{eqn:l}, Eq.~\ref{eqn:t}$
\STATE $\hat{y}^{s(i)} \leftarrow Eq.~\ref{eqn:y}$
\STATE record $\tau_{1:s}^{(i)}$
\ENDFOR
% \STATE $\hat{y}^{(i)}\leftarrow Eq.~\ref{eqn:y}$
\STATE $R^{(i)}\leftarrow Eq.~\ref{eqn:R}$
\ENDFOR
\STATE $\hat{y} = \frac{1}{M}\sum_{i=1}^{M}\hat{y}^{S(i)}$
\STATE $L_c,\nabla_\Theta \bar{R}\leftarrow Eq.~\ref{eqn:L_c},Eq.~\ref{eqn:delta L_R}$
\STATE $\Theta \leftarrow \Theta - \nabla_\Theta L_c + \nabla_\Theta \bar{R}$

\ENDWHILE

\RETURN $\Theta$

\end{algorithmic}

\end{algorithm}

\section{Experiments}
\begin{table*}[!ht]
\centering
\caption{The prediction performance of the proposed approach and other state-of-the-art methods. * denotes attention based state-of-the-art. The best performance is indicated in bold.}
\label{tab:comparison}
\scalebox{0.95}{
\begin{tabular}{c|cccccccc}
\hline
\multirow{5}{*}{MH} & Method    & MC-CNN      & C-Fusion    & MARCEL      & E-LSTM      & PRCA*        & WAS-LSTM*   & \textbf{Ours*}       \\ \cline{2-9} 
                         & Accuracy  & 87.19$\pm$0.77 & 88.66$\pm$0.62 & 92.35$\pm$0.46 & 91.58$\pm$0.38 & 93.32$\pm$0.75 & 91.42$\pm$1.25 & \textbf{96.12$\pm$0.37} \\
                         & Precision & 86.50$\pm$0.61 & 86.36$\pm$0.72 & 93.17$\pm$0.84 & 90.50$\pm$0.68 & 92.11$\pm$0.96  & 91.35$\pm$0.70 & \textbf{95.46$\pm$0.33} \\
                         & Recall    & 87.29$\pm$0.44 & 89.68$\pm$0.72 & 92.81$\pm$0.44 & 91.58$\pm$0.59 & 92.25$\pm$0.94  & 91.99$\pm$1.04 & \textbf{96.76$\pm$0.30} \\
                         & F1        & 86.89$\pm$0.66 & 87.98$\pm$0.79 & 92.98$\pm$0.74 & 91.03$\pm$0.68 & 92.17$\pm$1.06  & 91.66$\pm$1.21 & \textbf{96.10$\pm$0.47} \\ \hline
\multirow{5}{*}{PMP}  & Method    & MC-CNN      & C-Fusion    & MARCEL      & E-LSTM      & PRCA*        & WAS-LSTM*   & \textbf{Ours*}       \\ \cline{2-9} 
                         & Accuracy  & 81.16$\pm$1.32 & 81.86$\pm$0.74 & 82.87$\pm$0.81 & 83.21$\pm$0.68 & 82.39$\pm$1.04  & 84.89$\pm$2.18 & \textbf{90.33$\pm$0.62} \\
                         & Precision & 81.57$\pm$0.89 & 81.63$\pm$0.53 & 83.51$\pm$0.71 & 84.01$\pm$0.54 & 82.44$\pm$0.99  & 84.44$\pm$1.54 & \textbf{89.25$\pm$0.78} \\
                         & Recall    & 81.43$\pm$0.64 & 81.96$\pm$0.89 & 81.12$\pm$0.79 & 83.88$\pm$0.74 & 82.86$\pm$0.90  & 84.20$\pm$1.83 & \textbf{90.49$\pm$0.94} \\
                         & F1        & 81.50$\pm$0.72 & 81.79$\pm$0.71 & 82.29$\pm$0.76 & 83.94$\pm$0.95 & 82.64$\pm$1.19  & 84.81$\pm$1.06 & \textbf{89.86$\pm$0.81} \\ \hline
\multirow{5}{*}{HAR} & Method    & MC-CNN      & C-Fusion    & MARCEL      & E-LSTM      & PRCA*        & WAS-LSTM*   & \textbf{Ours*}       \\ \cline{2-9} 
                         & Accuracy  & 75.86$\pm$0.59 & 74.64$\pm$0.78 & 80.16$\pm$0.72 & 80.78$\pm$0.94 & 81.29$\pm$1.22  & 71.29$\pm$1.08 & \textbf{85.72$\pm$0.83} \\
                         & Precision & 76.93$\pm$0.78 & 73.30$\pm$0.75 & 81.63$\pm$0.50 & 81.34$\pm$0.43 & 80.55$\pm$1.26  & 70.76$\pm$0.93 & \textbf{85.61+0.53}  \\
                         & Recall    & 75.81$\pm$0.39 & 74.07$\pm$0.48 & 80.81$\pm$0.64 & 80.63$\pm$0.54 & 81.66$\pm$1.03  & 71.10$\pm$1.37 & \textbf{85.08$\pm$0.72} \\
                         & F1        & 76.36$\pm$1.11 & 73.68$\pm$0.79 & 81.21$\pm$0.85 & 80.98$\pm$0.64 & 81.11$\pm$1.02  & 70.92$\pm$1.16 & \textbf{85.34$\pm$0.58} \\ \hline
\multirow{5}{*}{MARS}    & Method    & MC-CNN      & C-Fusion    & MARCEL      & E-LSTM      & PRCA*        & WAS-LSTM*   & \textbf{Ours*}       \\ \cline{2-9} 
                         & Accuracy  & 81.34$\pm$0.59 & 81.48$\pm$0.56 & 81.68$\pm$0.87 & 81.59$\pm$0.77 & 85.38$\pm$0.82  & 74.82$\pm$1.42 & \textbf{88.29$\pm$0.87} \\
                         & Precision & 81.68$\pm$0.62 & 81.84$\pm$0.68 & 81.23$\pm$0.84 & 81.79$\pm$0.85 & 85.99$\pm$1.07  & 75.89$\pm$1.54 & \textbf{88.75$\pm$0.81} \\
                         & Recall    & 81.06$\pm$0.90 & 82.15$\pm$0.82 & 82.44$\pm$0.54 & 81.65$\pm$0.93 & 84.95$\pm$0.95  & 74.80$\pm$1.63 & \textbf{87.20$\pm$0.67} \\
                         & F1        & 81.32$\pm$0.42 & 81.99$\pm$0.64 & 81.85$\pm$0.97 & 81.71$\pm$0.81 & 85.46$\pm$1.02  & 75.34$\pm$1.27 & \textbf{87.96$\pm$0.74} \\ \hline
\end{tabular}
}
\end{table*}

\begin{table*}[!ht]
\centering
%\begin{scriptsize}
\caption{Ablation Study. S1 $\sim$ S6 are six structures by systematically removing five components from the proposed model. 
The considered components are: a) the selection module, (b) the partial observation processing from $e^s_1$ to $o^s_i$ ($i \in \{1,2,3\}$), (c) the convolutional merge of shared observations, (d) the temporal attentive selection (e) the multi-agent for selection.}
\label{tab:ablation}
\scalebox{0.95}{
\begin{tabular}{c|cccc|cccc}
\hline
\multirow{2}{*}{Ablation} & \multicolumn{4}{c|}{MHEALTH}                                                              & \multicolumn{4}{c}{PAMAP2}                                                                \\ \cline{2-9} 
                          & Accuracy             & Precision            & Recall               & F1                   & Accuracy             & Precision            & Recall               & F1                   \\ \hline
S1                        & 85.75$\pm$0.70          & 84.67$\pm$0.92          & 85.83$\pm$0.78          & 84.34$\pm$0.89          & 79.60$\pm$0.56          & 79.86$\pm$0.51          & 79.68$\pm$0.49          & 79.57$\pm$0.83          \\
S2                        & 80.59$\pm$1.54          & 80.37$\pm$0.95          & 80.95$\pm$1.44          & 80.00$\pm$1.12          & 71.49$\pm$1.55          & 71.62$\pm$1.18          & 71.38$\pm$1.36          & 71.49$\pm$1.42          \\
S3                        & 85.49$\pm$0.88          & 85.67$\pm$0.35          & 84.62$\pm$0.71          & 85.14$\pm$0.86          & 77.68$\pm$0.73          & 77.35$\pm$0.52          & 77.74$\pm$0.82          & 77.04$\pm$0.59          \\
S4                        & 88.32$\pm$0.75          & 87.11$\pm$0.96          & 87.25$\pm$0.94          & 87.17$\pm$1.06          & 78.39$\pm$1.04          & 78.44$\pm$0.99          & 78.86$\pm$0.90          & 78.64$\pm$1.19          \\
S5                        & 91.93$\pm$0.94          & 90.85$\pm$0.85          & 91.35$\pm$0.73          & 91.88$\pm$0.81          & 83.53$\pm$0.95          & 83.66$\pm$0.85          & 83.38$\pm$0.61          & 83.51$\pm$0.74          \\
\textbf{S6}               & \textbf{96.12$\pm$0.37} & \textbf{95.46$\pm$0.33} & \textbf{96.76$\pm$0.30} & \textbf{96.10$\pm$0.47} & \textbf{90.33$\pm$0.62} & \textbf{89.25$\pm$0.78} & \textbf{90.49$\pm$0.94} & \textbf{89.86$\pm$0.81} \\ \hline
\multirow{2}{*}{Ablation} & \multicolumn{4}{c|}{UCI HAR}                                                              & \multicolumn{4}{c}{MARS}                                                                  \\ \cline{2-9} 
                          & Accuracy             & Precision            & Recall               & F1                   & Accuracy             & Precision            & Recall               & F1                   \\ \hline
S1                        & 73.68$\pm$0.73          & 73.79$\pm$0.54          & 73.98$\pm$0.59          & 73.38$\pm$0.42          & 79.78$\pm$0.89          & 79.37$\pm$0.82          & 79.49$\pm$0.42          & 79.31$\pm$0.64          \\
S2                        & 68.95$\pm$2.85          & 67.41$\pm$2.97          & 67.92$\pm$2.88          & 67.66$\pm$2.81          & 71.94$\pm$2.18          & 71.34$\pm$2.52          & 71.99$\pm$2.67          & 71.15$\pm$2.71          \\
S3                        & 75.45$\pm$0.77          & 75.54$\pm$1.27          & 75.49$\pm$0.88          & 75.88$\pm$0.94          & 78.34$\pm$0.84          & 78.42$\pm$0.83          & 78.48$\pm$0.98          & 78.40$\pm$0.99          \\
S4                        & 76.29$\pm$1.22          & 76.55$\pm$1.26          & 77.66$\pm$1.03          & 77.11$\pm$1.02          & 81.38$\pm$0.82          & 81.99$\pm$1.07          & 81.95$\pm$0.95          & 81.46$\pm$1.02          \\
S5                        & 80.71$\pm$0.94          & 80.95$\pm$1.82          & 80.44$\pm$0.74          & 80.11$\pm$0.52          & 85.51$\pm$0.86          & 84.94$\pm$0.73          & 85.73$\pm$0.66          & 84.81$\pm$0.98          \\
\textbf{S6}               & \textbf{85.72$\pm$0.83} & \textbf{85.61+0.53}  & \textbf{85.08$\pm$0.72} & \textbf{85.34$\pm$0.58} & \textbf{88.29$\pm$0.87} & \textbf{88.75$\pm$0.81} & \textbf{87.20$\pm$0.67} & \textbf{87.96$\pm$0.74} \\ \hline
\end{tabular}
}
\end{table*}

\begin{figure*}[!ht]
\centering
\subfloat[Standing $l_1$]{
        \includegraphics[width=1.1in]{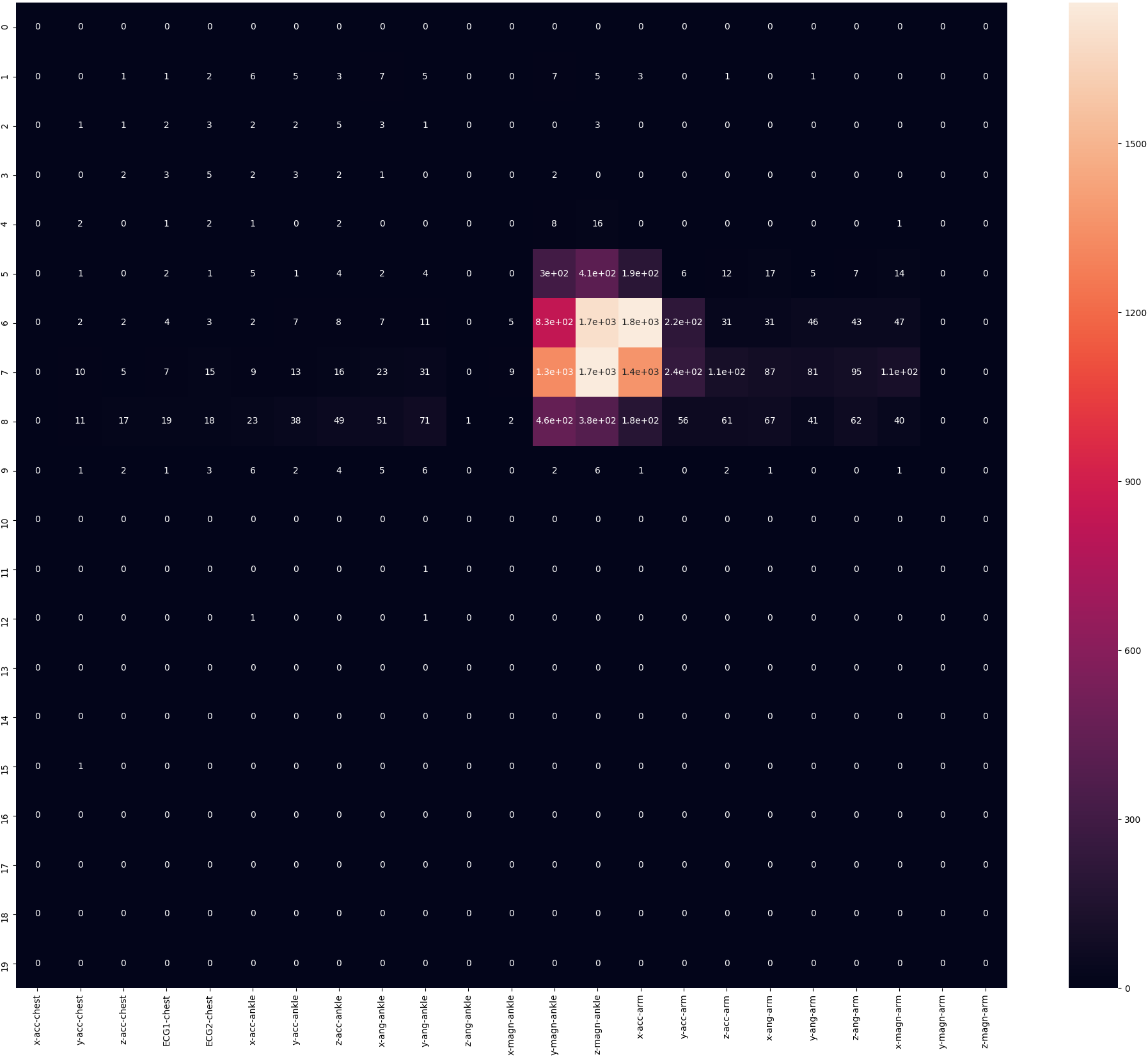}}
\subfloat[Standing $l_2$]{
        \includegraphics[width=1.1in]{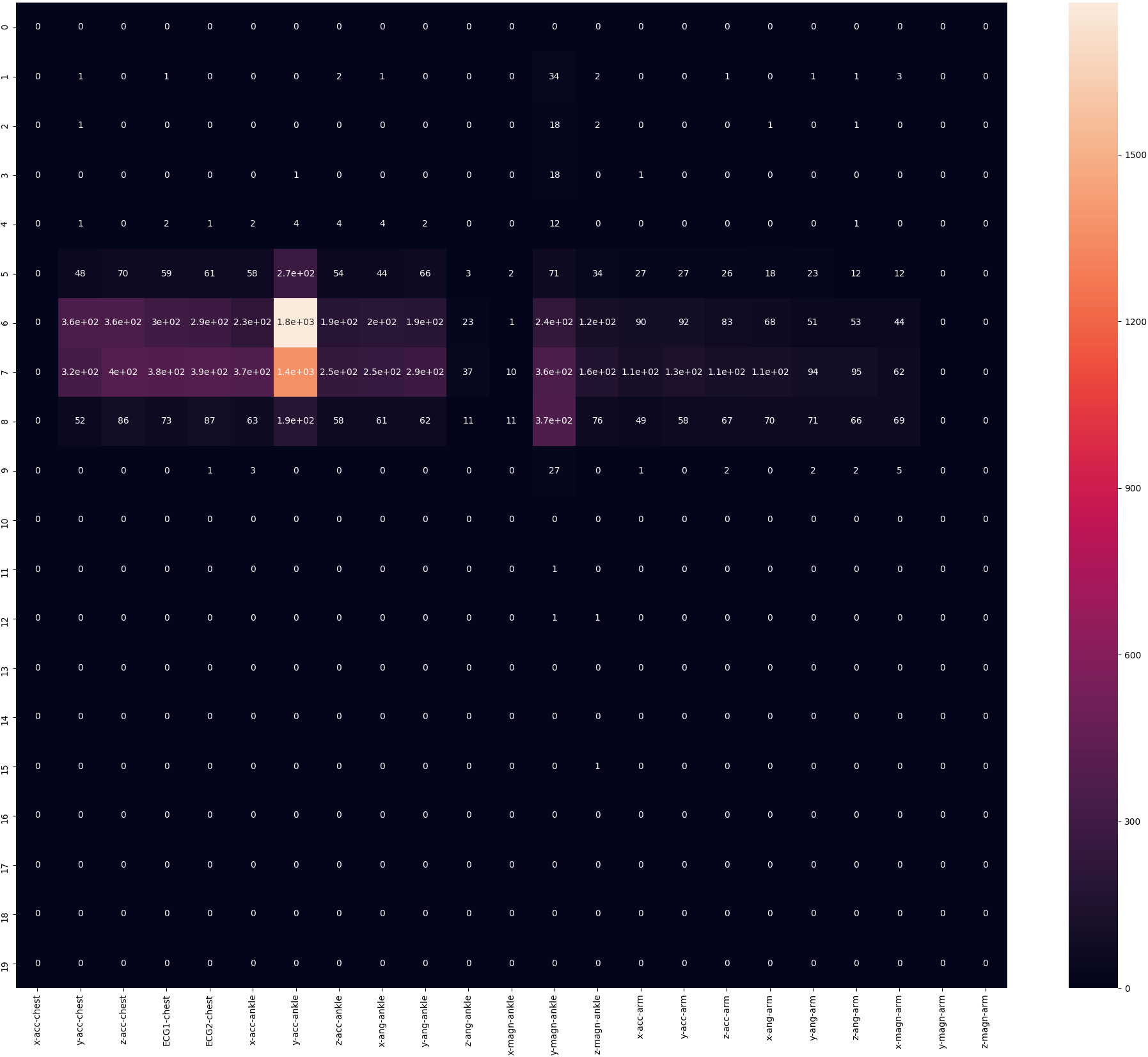}}
\subfloat[Standing $l_3$]{
        \includegraphics[width=1.1in]{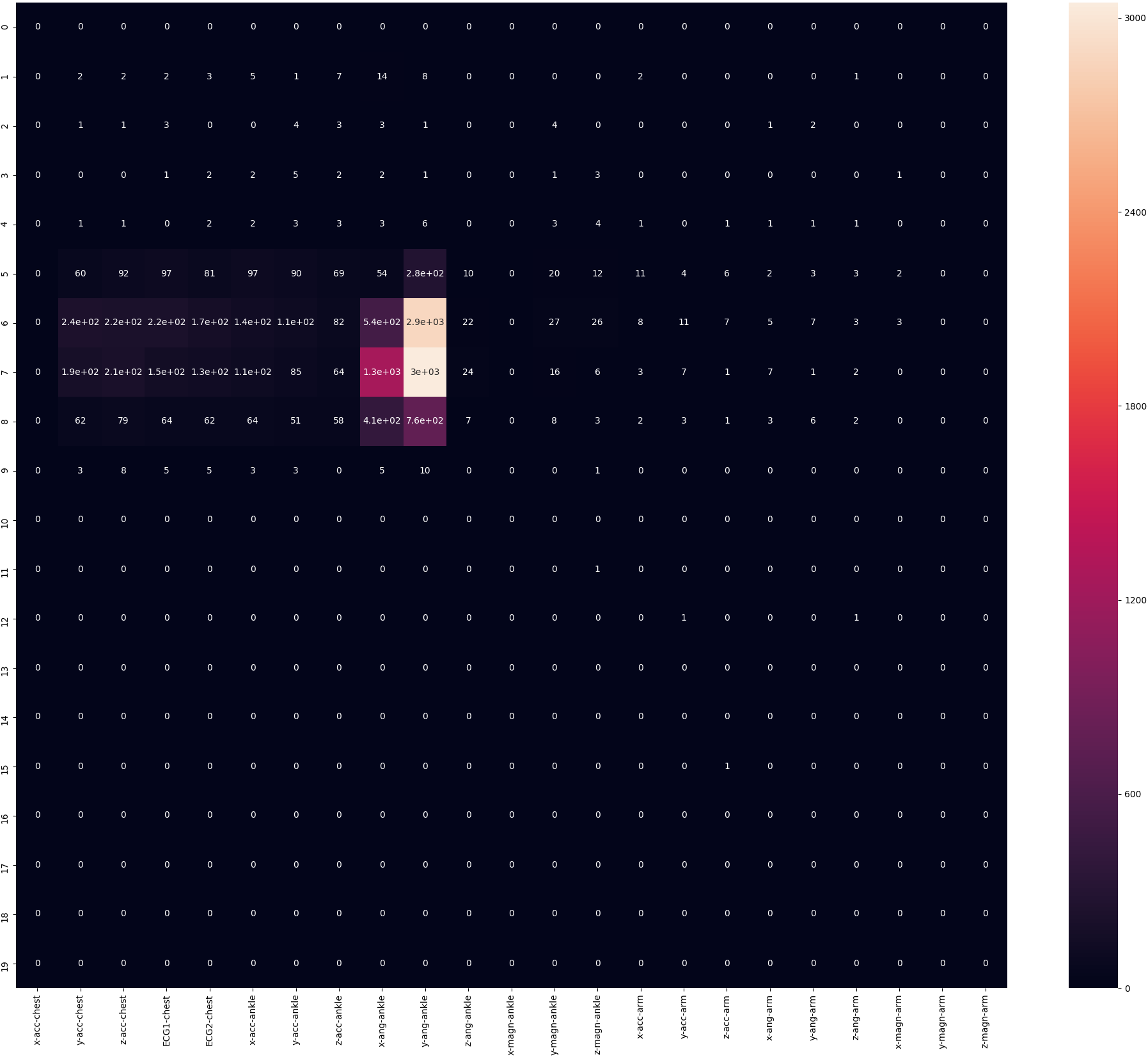}}
\subfloat[Going Upstairs $l_1$]{
        \includegraphics[width=1.1in]{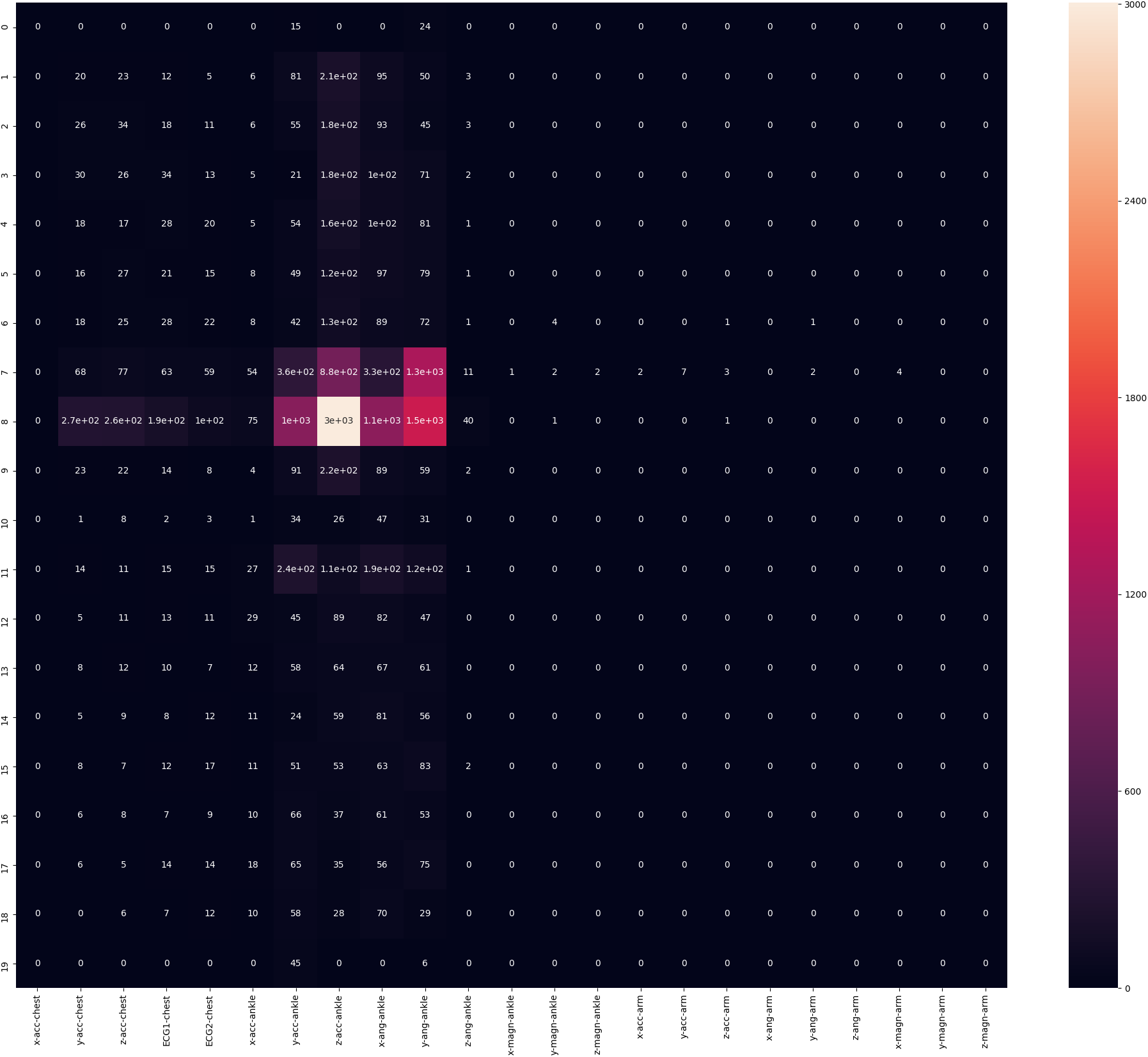}}
\subfloat[Going Upstairs $l_2$]{
        \includegraphics[width=1.1in]{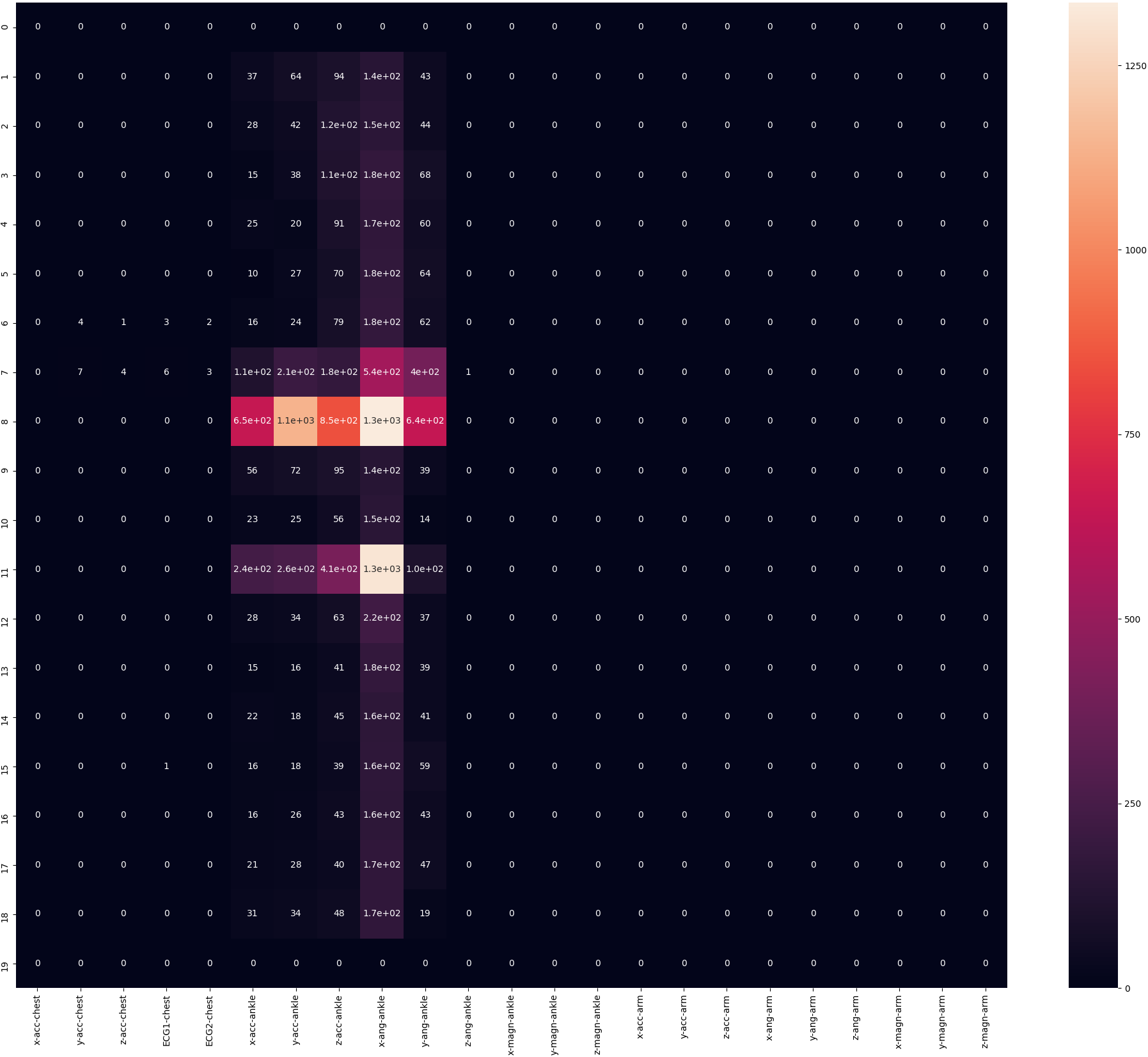}}  
\subfloat[Going Upstairs $l_3$]{
        \includegraphics[width=1.1in]{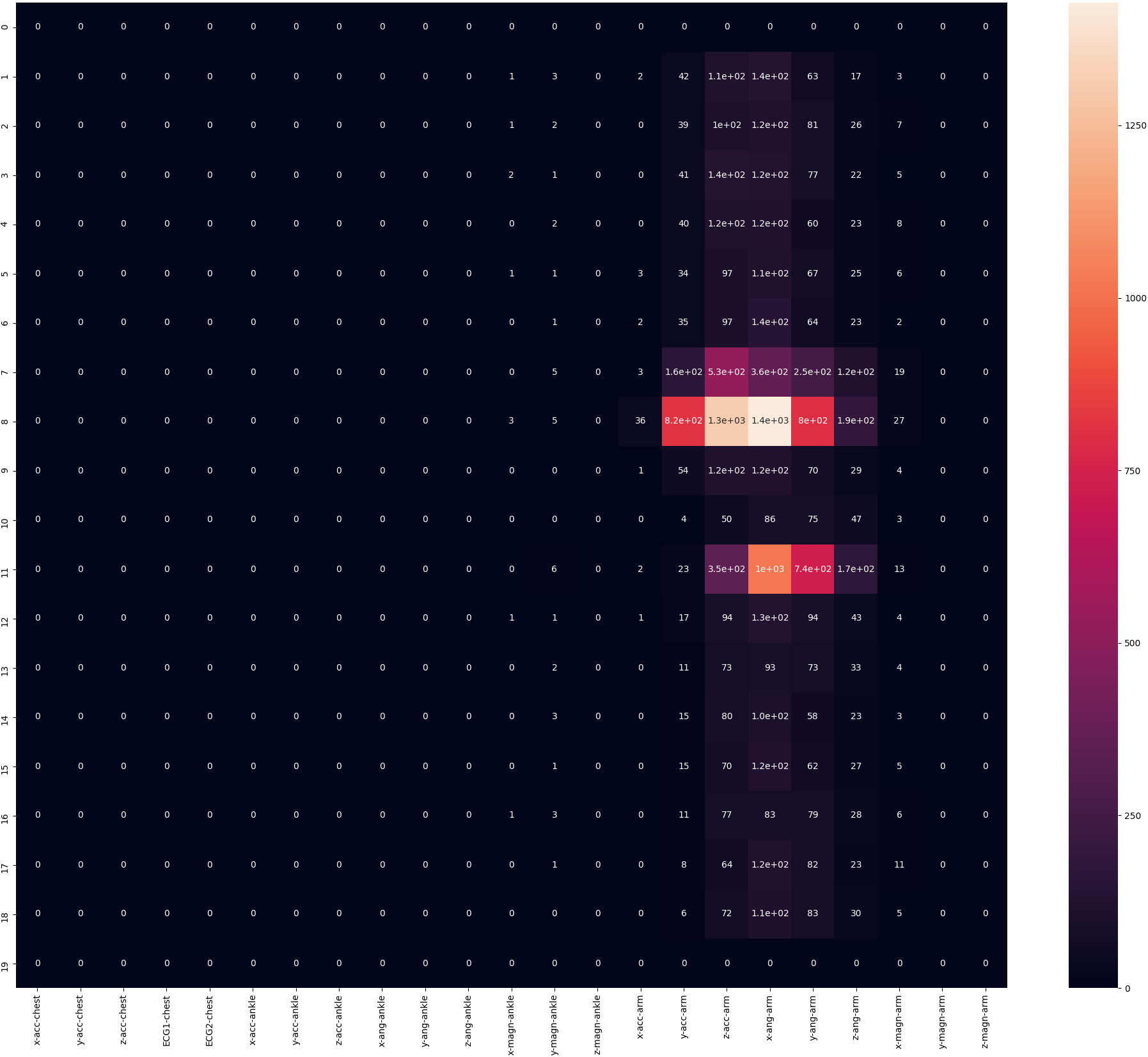}}        
        
\subfloat[Running $l_1$]{
        \includegraphics[width=1.1in]{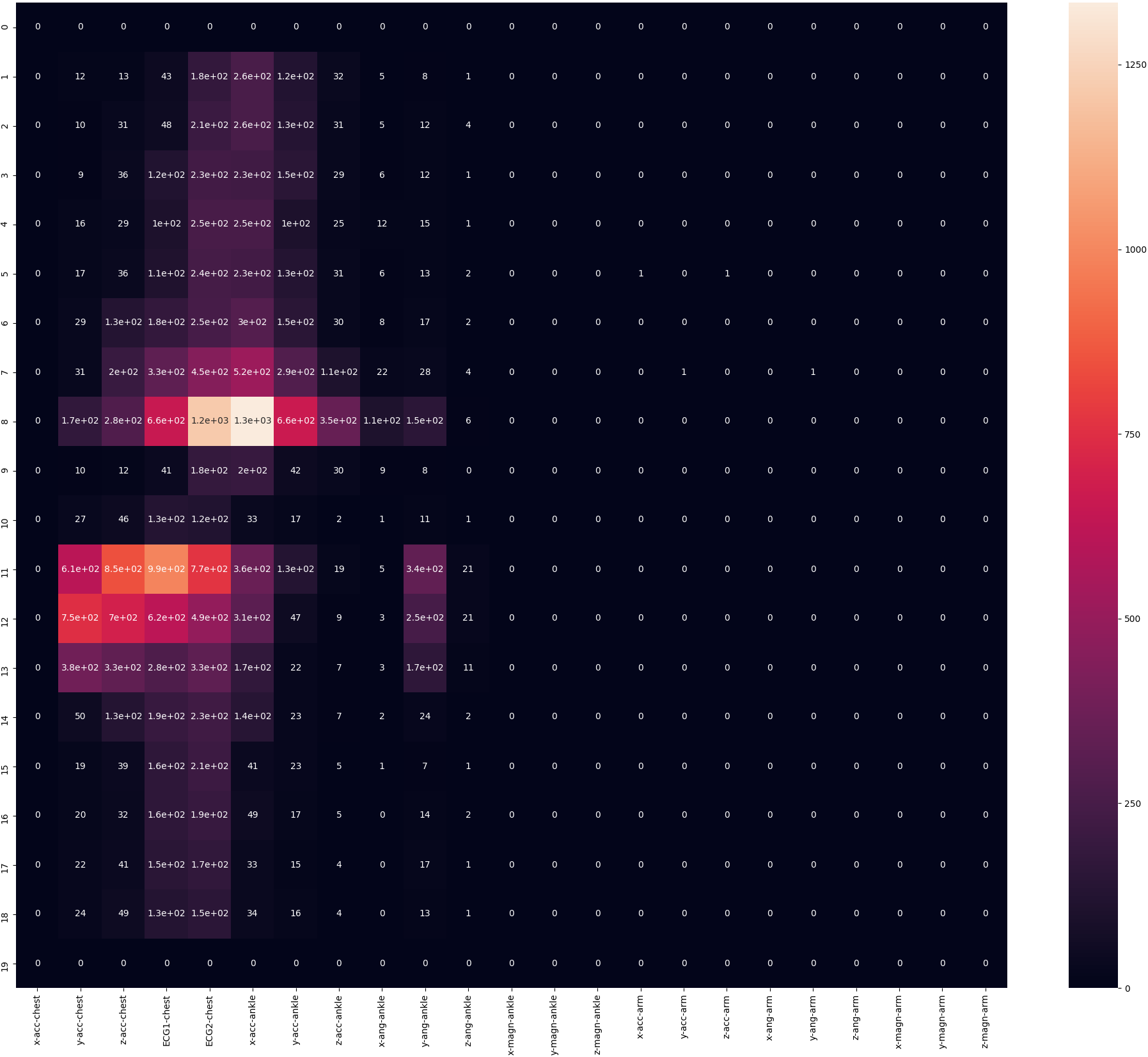}}               
\subfloat[Running $l_2$]{
        \includegraphics[width=1.1in]{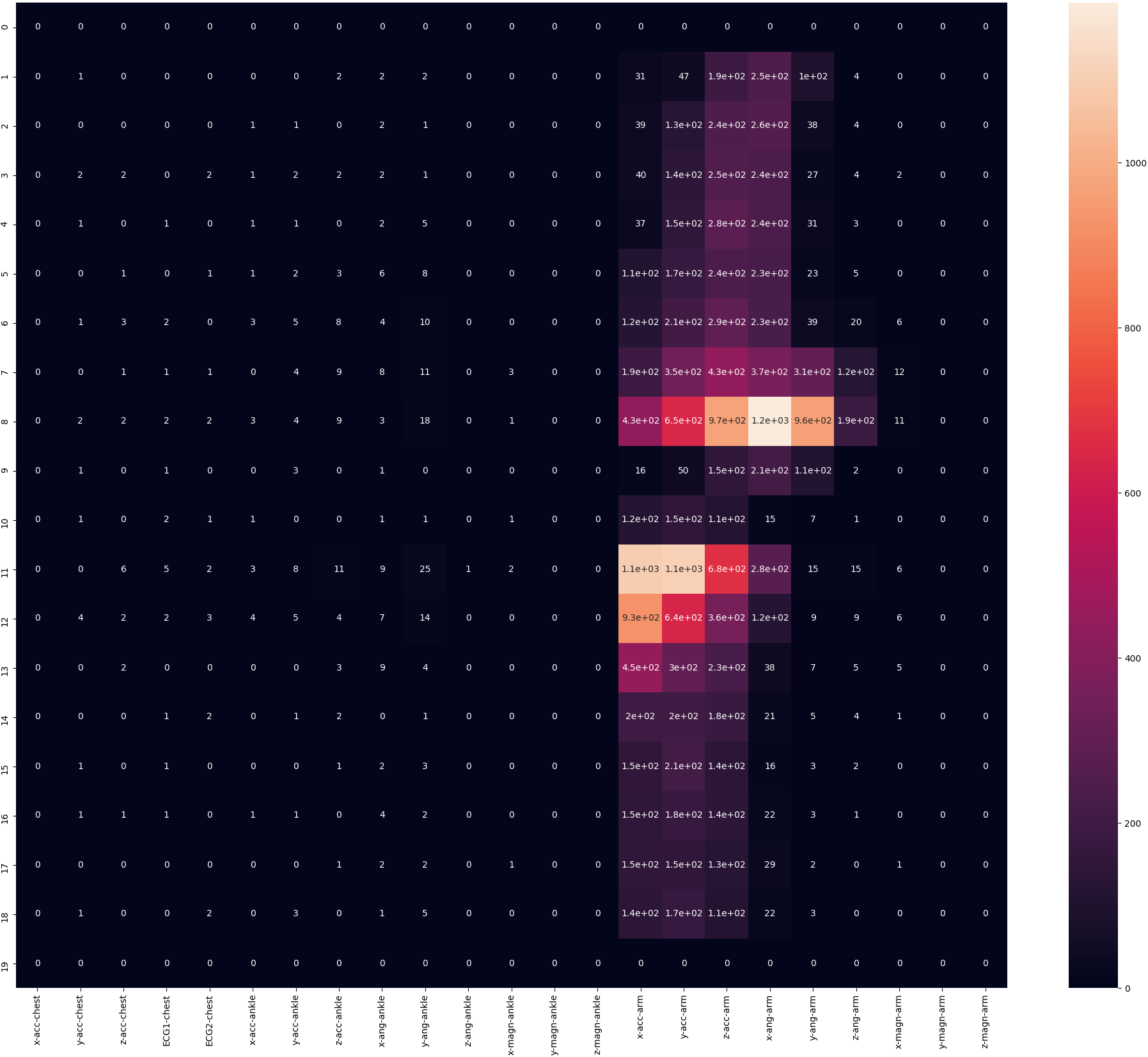}}  
\subfloat[Running $l_3$]{
        \includegraphics[width=1.1in]{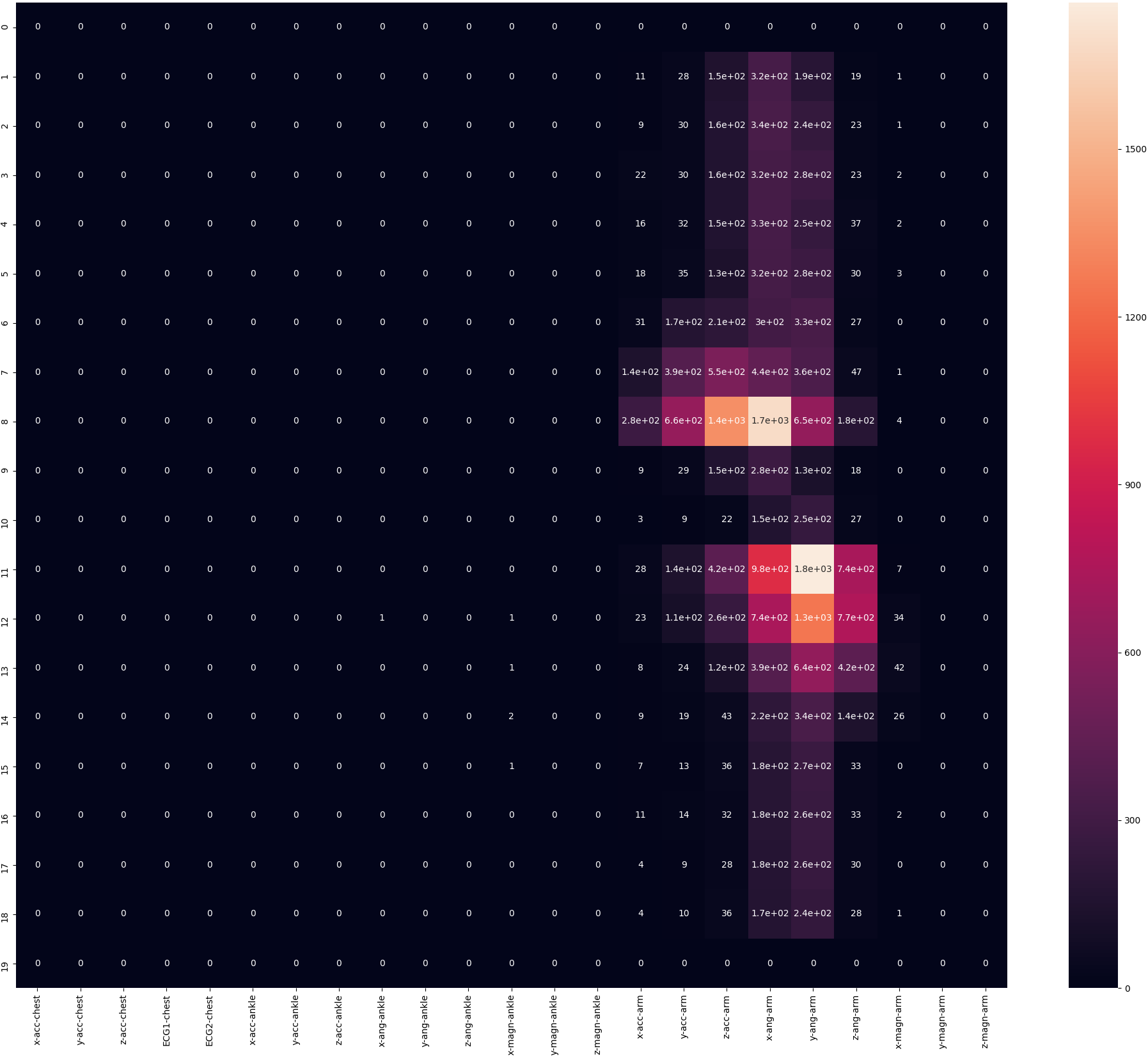}} 
\subfloat[Lying $l_1$]{
        \includegraphics[width=1.1in]{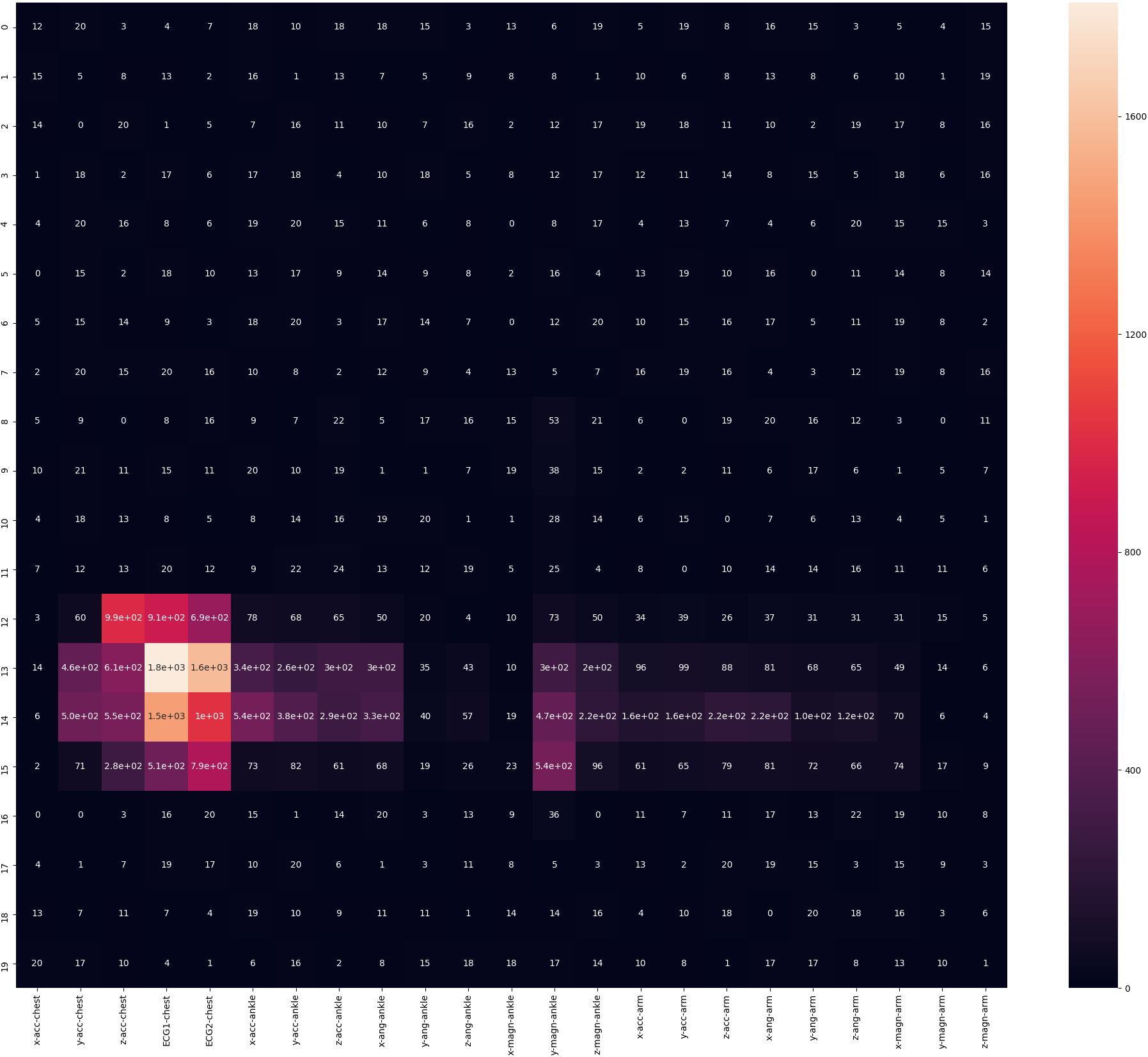}}
\subfloat[Lying $l_2$]{
        \includegraphics[width=1.1in]{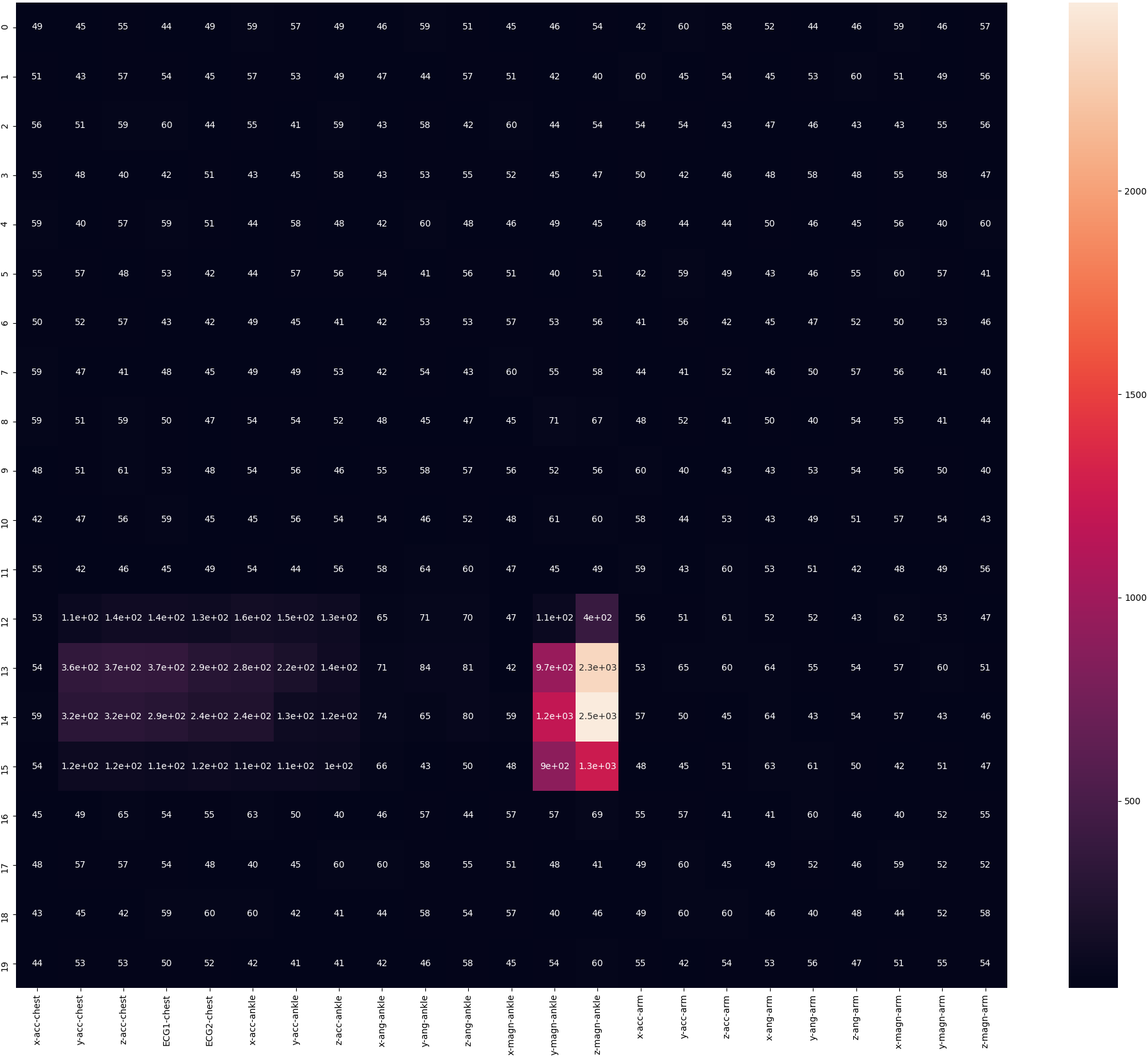}}
\subfloat[Lying $l_3$]{
        \includegraphics[width=1.1in]{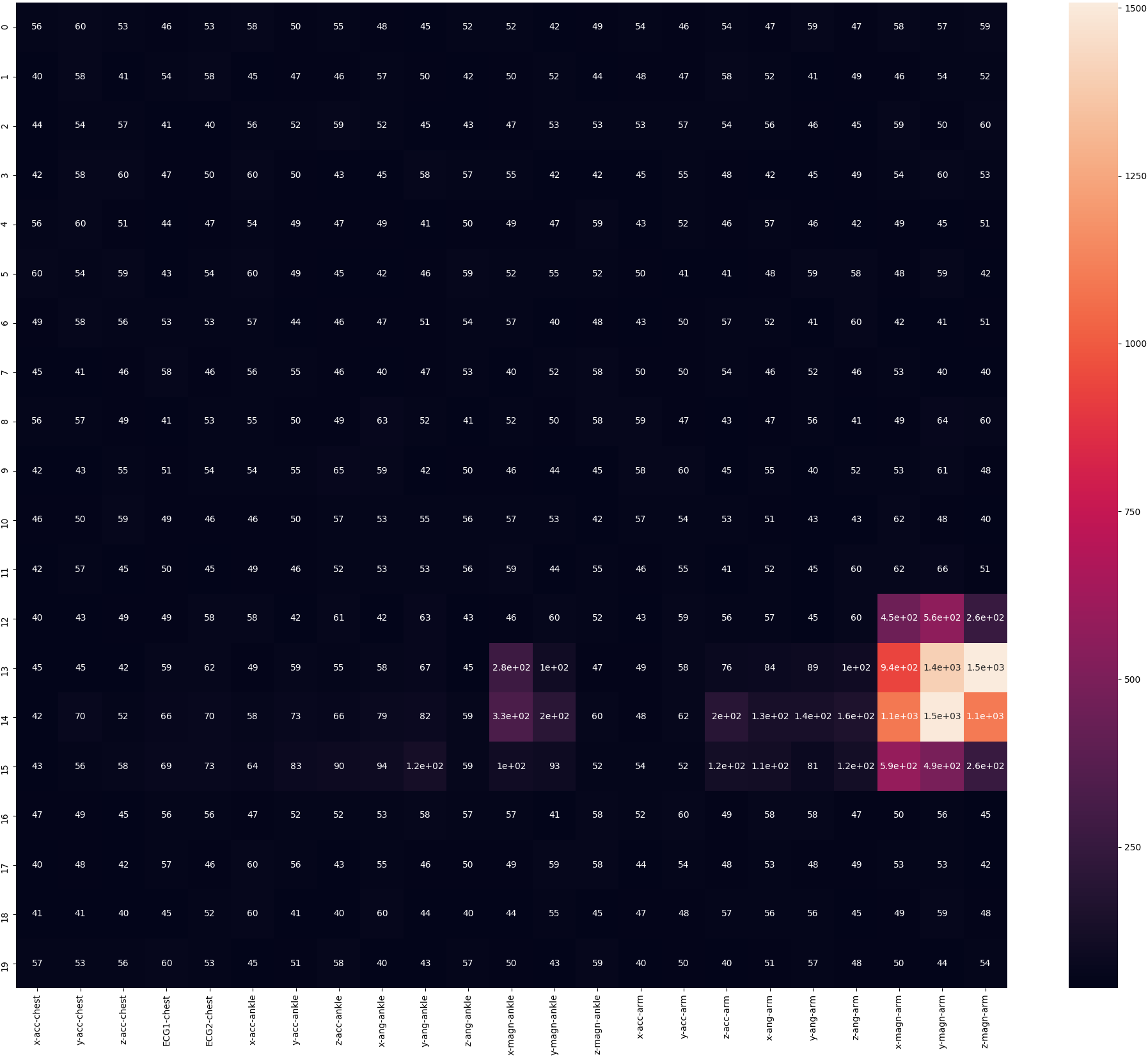}} 

\caption{Visualization of the selected modalities and time on MHEALTH. The input matrices' size is $20\times 23$, where $20$ is the length of the time window and $23$ is the number of modalities. Thus each grid denotes an input feature, and the values in the grids represent the frequency with which this feature is selected. Lighter colors denote higher frequency. To be clear, detailed illustration is provided in Table~\ref{tab:location}.}
\label{fig:visual}
\end{figure*}

\subsection{Experiment Setting}

We now introduce the settings in our experiments.
%\textcolor{blue}{We use ... datasets for evaluating....add one sentence for datasets} 
The time window of inputs is $20$ with $50\%$ overlap.
The size of each observation patch is set to $\frac{K}{8}\times\frac{P}{8}$, where $K\times P$ is the size of the inputs. 
In the partial observation part, the sizes of $\theta_e,\theta_{tl}, \theta_o$ are $128, 128, 220$, respectively. The filter size of the convolutional layer in the shared observation module is $1$ $\times$ $M$ and the number of feature maps is $40$, where $M$ denotes the width of $o_g^s$. The size of LSTM cells is $220$, and the length of episodes is $40$. The Gaussian distribution that defines the selection policies is with a variance of $0.22$. 
% When implementing our model, the computational cost is only $3\times\frac{1}{8}\times\frac{1}{8}$ of conventional deep learning models, and
% % $Cost= O(CNN) + O(LSTM) = O(3\times\frac{1}{8}\times\frac{1}{8}\times K\times 1\times1\times P\times1\times40))
% % +O(3\times\frac{K}{8}\times\frac{1}{8}\times40\times220\times40\times(4+12))
% % = O(264000\times K\times P)$, where $12$ is the number of label classes.
% the number of parameters is: $S(K,P) = 
% % S(Partial Observation)+S(Shared Observation) + S(LSTM) +S(Action) =
% O(3\times(\frac{K}{8}\times\frac{P}{8}\times128 + 2\times128+128\times220))
% +O(1\times3\times1\times40+220\times1\times1\times40)
% +O(4\times(220\times220\times1\times40+220\times220+220))
% +O(220\times(4+12))
% =O(6KP+2millions)
% =O(n)$.

% The classification is calculated by the softmax functions. 
% Each linear or convolutional layer is followed by a rectified linear unit (ReLU) activation.
% Dropout regularization with a keep probability of $0.5$ is applied before the fully-connected layers.
% We use stochastic gradient descent with Adam update rule to minimize the loss functions.
% The network parameters are optimized with a linearly annealed learning rate from $10^{-3}$ to $10^{-5}$. 
% All the experiments are conducted on a Nvidia Titan X Pascal GPU.

To ensure the rigorousness, the experiments are performed by Leave-One-Subject-Out (LOSO)
% cross-validation 
on four datasets, MHEALTH \cite{banos2014mhealthdroid}, PAMAP2 \cite{reiss2012introducing},
UCI HAR \cite{anguita2013public} and MARS. They contain $10,9,30,8$ subjects' data, respectively. 
% LOSO requires that the model should be tested on one subject's data and trained on the rest. We repeat the process, each time leaving out a different subject to use as the single test case, and average the results.
% The test set contains one subject's data and the training set contains the rest. We test all the subjects and take the average as the final result. 

\subsection{Comparison with State-of-the-Art}

To verify the overall performance of the proposed model, we first compare our model with other state-of-the-art methods. The compared methods include a convolutional model on multichannel time series for HAR (MC-CNN) \cite{yang2015deep}, a CNN-based multi-modal fusion model (C-Fusion) \cite{radu2018multimodal}, a deep multimodal HAR model with classifier ensemble (MARCEL) \cite{guo2016wearable}, an ensemble of deep LSTM learners for activity recognition (E-LSTM) \cite{guan2017ensembles}, a parallel recurrent model with convolutional attentions (PRCA) \cite{chen2018interpretable} and a weighted average spatial LSTM with selective attention (WAS-LSTM) \cite{zhang2018multi}.

As can be observed in Table~\ref{tab:comparison}, with respect to the datasets, MARCEL, E-LSTM, PRCA, WAS-LSTM and the proposed model perform better than MC-CNN and C-Fusion in MHEALTH and PAMAP2, as these models enjoy higher variance. They fit well when data contain numerous features and complex patterns. On the other hand, data in UCI HAR and MARS have fewer features, but MARCEL, E-LSTM, PRCA and our model still perform well while the performance of WAS-LSTM deteriorates. The reason is that WAS-LSTM is based on a complex structure and it requires more features as input.
In contrast, MARCEL and E-LSTM adopt rather simple models like DNNs and LSTMs. Despite the ensembles, they are still suitable for fewer features. PRCA and the proposed model select salient features directly with intuitive rewards, so they do not necessarily need a large number of features as well. 
In addition, the attention based methods, PRCA and WAS-LSTM, are more unstable than the other methods since the selection is stochastic and they cannot guarantee the effectiveness of all the selected features.
Overall, our model outperforms the compared state-of-the-art and eliminates the instability of regular selective attentions.

\subsection{Ablation Study}

We perform a detailed ablation study to examine the contributions of the proposed model components to the prediction performance in Table~\ref{tab:ablation}. Considering that there are five removable components in this model: (a) the modality selection module, (b) the transformation from $e_i^s$ to $o_i^s$ ($i \in \{1,2,3\}$), (c) the convolutional network for higher-level representations, (d) the temporal attentive selection (e) the multi-agent.  We consider six structures:
\textbf{S1}: We first remove the selection module including the observations, episodes, selections and rewards. For comparison, we set S1 to be a regular CNN as a baseline.
\textbf{S2}: We employ one agent but remove (b), (c), (d) and (e), so the workflow is: inputs $\rightarrow$ $e_i^s$ $\rightarrow$ LSTM $\rightarrow$ selections and rewards. The performance decreases and is more unstable than other structures. Although S2 includes attentions, the model does not include the previous selections in their observations, which influences their next decisions significantly.
\textbf{S3}: Based on S2, we add (b) to (a). (b) contributes considerably since the prediction results are improved by $5\%$ to $7\%$, because it feeds back the history selections to the agents for learning.
\textbf{S4}: We further consider (c) in the model. It can be observed that this setting also achieves better performance than S3 since it convolutionally merges the partial observations.
\textbf{S5}: (d) is added. The workflow is the same as S3, but the agents make an additional action: selecting $t^S$, which leads to another attention mechanism in time level. The performance is improved by $3\%$ to $5\%$.
\textbf{S6}: The proposed model. When combining all these benefits, our model achieves the best performance, higher than S5 by $5\%$ to $7\%$.

\subsection{Visualization and Explainability}
\begin{table}[ht]
\caption{The active modalities for activities selected by the agents are listed. $X, Y, Z$ denote the axis of data. Acc, Ang and Magn denote acceleration, angular velocity and magnetism, respectively. The most frequently selected locations are indicated in bold.}
\label{tab:location}
\begin{tabular}{ccc}
\hline
Activity                                                                                       & Agent                  & Location                                                                              \\ \hline
\multicolumn{1}{c|}{\multirow{3}{*}{Standing}}                                                 & \multicolumn{1}{c|}{1} & Y,\textbf{Z}-Magn-Ankle, X,Y-Acc-Arm                                                           \\ \cline{2-3} 
\multicolumn{1}{c|}{}                                                                          & \multicolumn{1}{c|}{2} & \textbf{Y}-Acc-Ankle, ECG1,2-Chest                                                             \\ \cline{2-3} 
\multicolumn{1}{c|}{}                                                                          & \multicolumn{1}{c|}{3} & X,\textbf{Y}-Ang-Ankle                                                                         \\ \hline
\multicolumn{1}{c|}{\multirow{3}{*}{\begin{tabular}[c]{@{}c@{}}Going\\ Upstairs\end{tabular}}} & \multicolumn{1}{c|}{1} & Y,\textbf{Z}-Acc-Ankle, X,Y-Ang-Ankle                                                          \\ \cline{2-3} 
\multicolumn{1}{c|}{}                                                                          & \multicolumn{1}{c|}{2} & X,Y,Z-Acc-Ankle, \textbf{X},Y-Ang-Ankle                                                        \\ \cline{2-3} 
\multicolumn{1}{c|}{}                                                                          & \multicolumn{1}{c|}{3} & Y,\textbf{Z}-Acc-Arm, X,Y,Z-Ang-Arm                                                            \\ \hline
\multicolumn{1}{c|}{\multirow{3}{*}{Running}}                                                  & \multicolumn{1}{c|}{1} & \begin{tabular}[c]{@{}c@{}}Y,Z-Acc-Chest, ECG1,2-Chest\\ \textbf{X},Y,Z-Acc-Ankle\end{tabular} \\ \cline{2-3} 
\multicolumn{1}{c|}{}                                                                          & \multicolumn{1}{c|}{2} & X,Y,Z-Acc-Arm, \textbf{X},Y-Ang-Arm                                                            \\ \cline{2-3} 
\multicolumn{1}{c|}{}                                                                          & \multicolumn{1}{c|}{3} & Y,Z-Acc-Arm, X,\textbf{Y}.Z-Ang-Arm                                                            \\ \hline
\multicolumn{1}{c|}{\multirow{3}{*}{Lying}}                                                    & \multicolumn{1}{c|}{1} & Y,Z-Acc-Chest, \textbf{ECG1},2-Chest                                                           \\ \cline{2-3} 
\multicolumn{1}{c|}{}                                                                          & \multicolumn{1}{c|}{2} & Y,\textbf{Z}-Magn-Ankle                                                                        \\ \cline{2-3} 
\multicolumn{1}{c|}{}                                                                          & \multicolumn{1}{c|}{3} & X,\textbf{Y},Z-Magn-Arm                                                                        \\ \hline
\end{tabular}
\end{table}

The proposed method decomposes the activities into participating motions, from each of which the agents decide the most salient modalities individually, which makes the model explainable. 
% For wearable sensor-based activity recognition, subjects usually wear more than one sensors on their dominant body parts like arms, chest, and ankles, each sensor with multimodal. Our model provides a superiority that it feeds the selected locations back at each step. 
% Based on the locations, it is easy to infer which body parts, even which modalities are selected as the salient features. 
We present the visualized process of recognizing standing, going upstairs, running and lying on MHEALTH. The available features include three $3$-axis acceleration from chests, arms and ankles, two ECG signals, two $3$-axis angular velocity and two $3$-axis magnetism vectors from arms and ankles.
Figure~\ref{fig:visual} shows the modality heatmaps of all agents. We observe that each agent does focus on only a part of modalities in a time period during recognition. Table~\ref{tab:location} lists the most frequently selected modalities. We can observe that magnetism (orientation) in standing and lying is selected as one of the most active features, owing to the fact it is easy to distinguish between standing and lying with people's orientation. 
Another example is that the most distinguishing characteristic of going upstairs is ``up". Therefore, Z-axis acceleration is specifically selected by agents for going upstairs. Also, identifying running involves acceleration, ECG, and arm swing, which conforms to the experiment evidence as well.
The agents also select several other features with lower frequencies, which avoids losing effective information. 
% With respect to time, it is noticeable only a short time period is particularly active. This demonstrates the temporal salience of the sensory data and the effectiveness of our model.
\section{Conclusion}
In this work, we first propose a selective attention method for spatially-temporally varying salience of features. Then, multi-agent is proposed to represent activities with collective motions. The agents' cooperate by aligning their actions to achieve their common recognition target.
We experimentally evaluate our model on four real-world datasets, and the results validate the contributions of the proposed model.
% The results demonstrate the outperformance of our model. 
% We detail an ablation of the components of the proposed model to demonstrate their contributions.
% Visualized results are also provided to present the selected locations, which shows the effectiveness and the explainability of the proposed model.

\bibliographystyle{named}
\bibliography{ijcai19}

\end{document}